\documentclass[conference]{IEEEtran}
\usepackage{graphicx}
\usepackage{subfigure}
\usepackage{xcolor}
\usepackage{amsmath,bm}

\usepackage{algorithm}
\usepackage{algorithmic}
\usepackage{multirow}
\usepackage{booktabs}
\usepackage{threeparttable}
\usepackage{flushend}

\newcommand{\algorithmickernelone}{\textbf{Kernel 1:}}
\newcommand{\KERNELONE}{\item[\algorithmickernelone]}
\newcommand{\algorithmickerneltwo}{\textbf{Kernel 2:}}
\newcommand{\KERNELTWO}{\item[\algorithmickerneltwo]}

\ifCLASSINFOpdf
\else
\fi
\begin{document}
%
\title{A Gb/s Parallel Block-based Viterbi Decoder for Convolutional Codes on GPU}

\author{\IEEEauthorblockN{Hao Peng, Rongke Liu, Yi Hou and Ling Zhao}
\IEEEauthorblockA{School of Electrical and Information Engineering\\
Beihang University,
Beijing, China\\
Email: \{haybla, rongke\_liu, mokyy, zhaoling\}@buaa.edu.cn}
}


%


\maketitle

\begin{abstract}
In this paper, 
we propose a parallel block-based Viterbi decoder (PBVD) on the graphic processing unit (GPU) platform for the decoding of convolutional codes. The decoding procedure is simplified and parallelized, and the characteristic of the trellis is exploited to reduce the metric computation. Based on the compute unified device architecture (CUDA), two kernels with different parallelism are designed to map two decoding phases. Moreover, the optimal design of data structures for several kinds of intermediate information are presented, to improve the efficiency of internal memory transactions. Experimental results demonstrate that the proposed decoder achieves high throughput of 598Mbps on NVIDIA GTX580 and 1802Mbps on GTX980 for the 64-state convolutional code, which are 1.5 times speedup compared to the existing fastest works on GPUs.
\end{abstract}


%
\IEEEpeerreviewmaketitle

\section{Introduction}


The convolutional codes are widely used in digital communication systems to correct the transmission errors, which have been adopted by almost all advanced wireless communication standards. There are strong requirements to develop high performance decoding components with good scalability and reconfigurability to support various standards, which can be applied to new radio communication systems such as the Software Defined Radio (SDR) and Cognitive Radio (CR). As the Viterbi algorithm \cite{Viterbi1967} is the most popular method for decoding convolutional codes, our discussion focuses on the techniques of Viterbi decoder implementations.

Traditional communication systems mainly use Field-Programmable Gate Array (FPGA) and Application Specific Integrated Circuit (ASIC) in hardware platforms. Enormous researches of the Viterbi decoder implementation are based on FPGA/ASIC and Gb/s throughput is achieved \cite{Fettweis1996} \cite{VLSI2010}. However, these high performances are always along with expensive cost and long development cycle, and these techniques can not provide the flexibility required by SDR or CR systems. Alternative microprocessors like Central Processor Unit (CPU) are more flexible than FPGAs/ASICs. Some works on CPU-based software decoding use single-instruction multiple-data (SIMD) instruction sets to achieve parallel decoding \cite{CPU2009} \cite{CPU2010}. But restricted by their computation resources, the data processing speed and decoding throughputs are much lower.


High performance computing (HPC) on GPUs is developed rapidly over the last decade.
Compared with FPGAs/ASICs, GPU-based implementations have very good flexibility and scalability using high-level programming language. Compared with CPUs, GPUs have more massive ALU cores to ensure large-scale parallel execution, which can gain higher throughput with appropriate optimization.
A lot more works have focused on GPU-based decoding in recent years. \cite{SDR2011} \cite{TVDA2011} \cite{TVDA_WCNC2013} \cite{SDR2010} \cite{TVDA_2014} use CUDA to design Viterbi decoders for SDR systems on NVIDIA GPUs. \cite{OPENCL2014} uses opening computing language (OpenCL) to achieve accelerating Viterbi decoding on an AMD GPU. However, most of these works just simply design a parallel decoder for block-coded convolutional codes, and basic level of optimizations are presented. Compared with these works, a better parallel Viterbi decoding algorithm with lower computational complexity is proposed in this paper. Fine-grained and coarse-grained parallelism optimizations are both presented, to maximize the execution efficiency of mathematical operations, memory transactions and data transfer between the host and the device. The good generality means our parallel block-based Viterbi decoder can work for most kinds of convolutional codes, and some optimizations can also be adopted to implementations of other GPU-based decoders.

\section{The Viterbi Decoding Algorithm}
The Viterbi algorithm (VA), a maximum likelihood sequence estimator, uses the trellis to exhaustively search the sequence that is closest to received bits from channel. It consists of two procedures in two directions: the forward procedure and the traceback procedure. Three kind of units will be calculated: the path metric (PM), the branch metric (BM) and the survivor path (SP). BM is calculated to measure Hamming/Euclidean distance from the received bits to the legal codewords at each stage. PM is the accumulated distance added by BMs. SP takes a record of the path with minimum distance to each state.
Forward computing starts at stage 0 with all metrics set to zero. For each state at current stage, an add-compare-select (ACS) operation is carried out to update their PM at next stage and rewrite their SP relatively. The ACS operation can be described by equation (\ref{Eq_ACS}). $PM_n^j$ denotes the path metric of state $j$ at stage $n$. $BM_n^{i,j}$ denotes the branch metric from state $i$ at stage $n-1$ to state $j$ at stage $n$. $PM_{n - 1}^i$ and $BM_n^{i,j}$ are added up for all state $i$ connected to state $j$ and a minimum result is chosen to update $PM_n^j$.
\begin{equation}\label{Eq_ACS}
PM_n^j = \mathop {\min }\limits_i \left( {PM_{n - 1}^i + BM_n^{i,j}} \right)
\end{equation}
While the forward ACS computing finishes at the end of the data stream, a state with minimum PM should be estimated as the beginning of traceback procedure. The selected state $S_E$ is believed to be the true encoding tail state with high probability. Therefore, the traceback process goes along the final survivor path $SP_T^E$ to obtain decoded bits.



In below sections, the $(R, 1, K)$ convolutional code with code rate $1/R$ and constraint length $K$ is concerned. The number of states is denoted by $N$.




\section{Proposed GPU-based Decoder and Methods for Efficient Decoding}

\subsection{Parallel Block-based Viterbi Decoder}
The original VA is not suitable to decode the convolutional codes encoded in a stream, as a huge amount of storage resource would be required and high decoding latency would not be acceptable. Thus, we propose a parallel block-based Viterbi decoder (PBVD) based on the GPU architecture.

\begin{figure}[b] 
\centering
\includegraphics[width=3.5in]{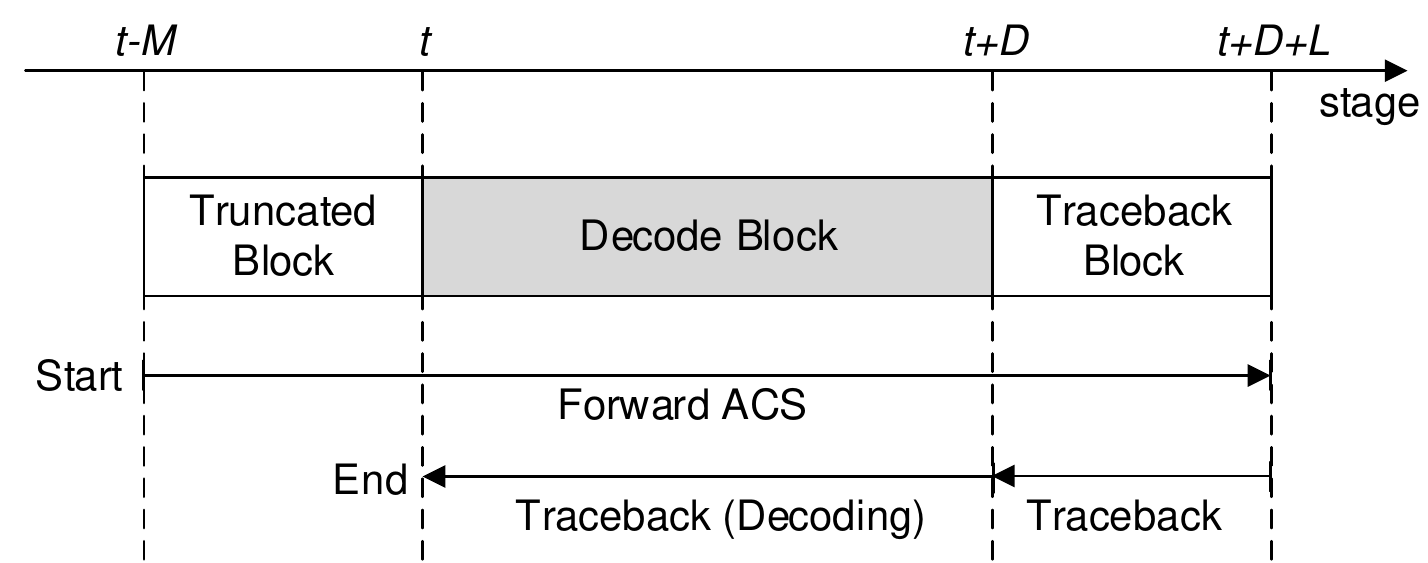}
\caption{The diagram of decoding procedures inside a data segment.}
\label{Fig_SBVD}
\end{figure}

Fig.\ref{Fig_SBVD} shows a schematic of the decoding procedures using PBVD. A real-time constraint is introduced into the decoding procedure. A data segment from stage $t-M$ to $t+D+L$ called a parallel block (PB) consists of a truncated block, a traceback block and a decoding block. Assuming that the block to be decoded starts at stage $t$, with the length of $D$, the PBVD should start at stage $t-M$. A forward ACS procedure is carried out with unknown initial state metrics (typically set to zero). The ACS operation goes through stage $t-M$ to $t+D+L$ and survivor paths with length of $M+D+L$ are estimated and stored, so as the PMs for each state. At the end of the interval, a traceback procedure starts from a random state (state $S_0$, for example). After $L$ times traceback along a randomly picked survivor path, state $S_E$ is reached and regarded as the authentic state at stage $t+D$. Afterwards, traceback procedure would continue and the data segment from stage $t$ to $t+D$ is decoded.

Unlike the original VA, there is no state estimation between forward and backward procedure in PBVD. That means the shortest path would not need to be picked out as the unique selection for backward decoding. This simplification benefits from the traceback block, which provides $L$ stage for all survivor paths merging to an authentic state at stage $t+D$. The length $L$ is called decoding depth and typically equal to $5K$ \cite{SBVD1997}. Similarly, by a number of $M$ iterations on the truncated block, the truncation error due to the unknown initial metrics is negligible. Thus, the strong probability of successful decoding for the mid block is guaranteed.

\begin{figure}[tb] 
\centering
\includegraphics[width=3.2in]{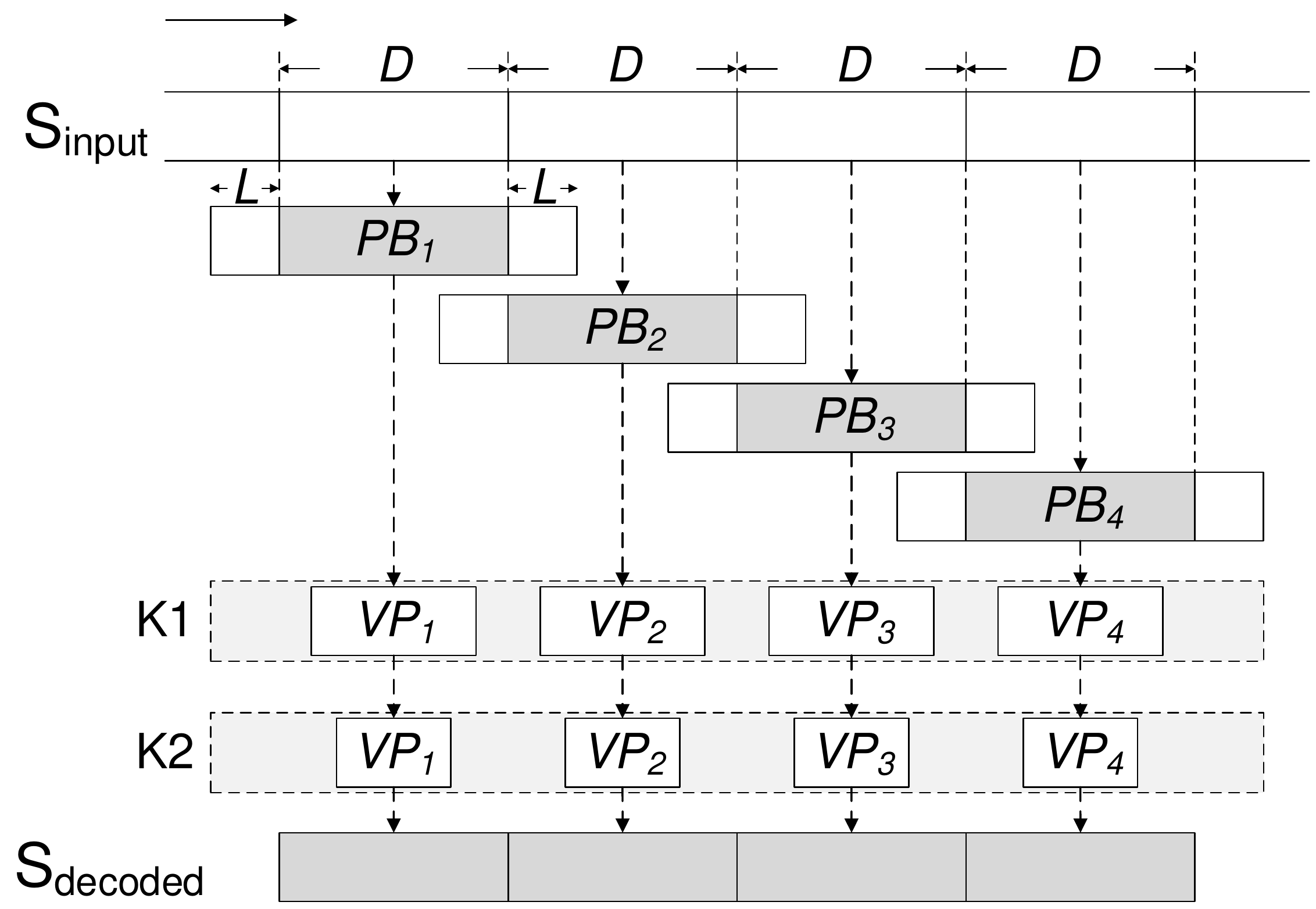}
\caption{The diagram of parallel decoding for data stream using two individual GPU kernels. (Note that the composition of VP in K1 and K2 are different.)}
\label{Fig_SBVDstream}
\end{figure}

To decode a stream of convolutional codes, the input data could be blocked to a series of segments of length $D$. Each segment extends a length of $L$ in both sides as the truncated block and traceback block, to form a parallel block ($M$ is set equal to $L$ in the following description), so the biting length for adjacent PB is $2L$.
To achieve high decoding throughput on a GPU, these $N_t$ PBs should be decoded concurrently by two individual GPU kernels (denoted by K1 and K2) with different parallelism to match the different computational complexity of procedures in two directions. Each PB should be successively handled by GPU thread cluster in K1 and K2, which are named virtual processors (VP). After synchronization, the outputs of all VPs in K2 are finally gathered to form the decoded stream. An example of the design for stream decoding using GPU-based PBVD with $N_t = 4$ is shown in Fig.\ref{Fig_SBVDstream}.






\subsection{Optimized Parallelism for Forward ACS Computation}
Typically, the commonly used schemes for the forward ACS operations are the state-based parallel execution \cite{TVDA_WCNC2013} and the butterfly-based parallel execution \cite{TVDA_2014}. In this paper, a group-based parallel scheme is proposed by exploiting the characteristics of the trellis, to reduce the amount of branch metric computation in the forward procedure.

For a $(R,1,K)$ convolutional code, the state in the trellis is defined by the contents of the $v$ binary memory cells $D_{v-1}{\sim}D_0$ in the encoder, which can be denoted by $S_d$ and $d=(D_{v-1}D_{v-2} \cdot \cdot \cdot D_1D_0)_2$. There are $R$ filters in the encoder, the $r$th of which has impulse response ${\textbf{\emph{g}}^{(r)}} = [ {g_{K - 1}^{(r)}g_{K - 2}^{(r)} \cdot  \cdot  \cdot g_1^{(r)}g_0^{(r)}} ]$, called the generator polynomials. $\textbf{\emph{c}}(S_d, x) = [ {{c^{(1)}}{c^{(2)}} \cdot  \cdot  \cdot {c^{(R)}}}]$ is used to express the encoder output corresponding to input bit $x$ at state $S_d$. $c^{(r)}$ is the output of the $r$th filter, 0 or 1, which can be calculated by:
\begin{equation}\label{Eq_cr}
{c^{(r)}} = (x \cdot g_{K - 1}^{(r)}) \oplus ({D_{K - 2}} \cdot g_{K - 2}^{(r)}) \oplus  \cdot  \cdot  \cdot  \oplus ({D_0} \cdot g_0^{(r)})\\
\end{equation}
All operations $\oplus$ are module-2 additions in field GF(2). Consider a butterfly structure from the trellis,
the contiguous states $S_{2j}$ and $S_{2j+1}$ in $j$th butterfly ($j = 0,1,2,...,N/2-1$) would like to shift to the states $S_j$ or $S_{j+2^{v-1}}$ for different input bits. $\bm{\alpha}$ and $\bm{\beta}$ denote the output of encoder at state $S_{2j}$ with input bit $x=0$ and 1 respectively. So as the $\bm{\gamma}$ and $\bm{\theta}$ for state $S_{2j+1}$. The $r$th bit ${\alpha}^{(r)}$ in $\bm{\alpha} = [ {{{\alpha}^{(1)}}{{\alpha}^{(2)}} \cdot  \cdot  \cdot {{\alpha}^{(R-1)}}{{\alpha}^{(R)}}}]$ can be obtained by:
\begin{align}\label{Eq_alpha}
{{\alpha}^{(r)}} &= c^{(r)}(S_{2j},0) \notag\\
&= (x \cdot g_{K - 1}^{(r)}) \oplus  \cdot  \cdot  \cdot  \oplus ({D_1} \cdot g_1^{(r)})  \oplus ({D_0} \cdot g_0^{(r)}) \notag\\
&= (0 \cdot g_{K - 1}^{(r)}) \oplus  \cdot  \cdot  \cdot  \oplus ({D_1} \cdot g_1^{(r)})  \oplus (0 \cdot g_0^{(r)}) \notag\\
&= ({D_{K - 2}} \cdot g_{K - 2}^{(r)}) \oplus  \cdot  \cdot  \cdot  \oplus ({D_1} \cdot g_1^{(r)})
\end{align}
Similarly, ${\beta}^{(r)}$, ${\gamma}^{(r)}$ and ${\theta}^{(r)}$ can be obtained as follows:
\begin{align}
\label{Eq_beta}
{{\beta}^{(r)}} &= c^{(r)}(S_{2j},1) = g_{K - 1}^{(r)} \oplus {\alpha}^{(r)} \\
\label{Eq_gamma}
{{\gamma}^{(r)}} &= c^{(r)}(S_{2j+1},0) = {\alpha}^{(r)} \oplus g_0^{(r)}\\
\label{Eq_theta}
{{\theta}^{(r)}} &= c^{(r)}(S_{2j+1},1) = g_{K - 1}^{(r)} \oplus {\alpha}^{(r)} \oplus g_0^{(r)}
\end{align}


From equation (\ref{Eq_alpha}) to (\ref{Eq_theta}) we can conclude that for given generator polynomials, once the $\bm{\alpha}$ is established, other outputs $\bm{\beta}$, $\bm{\gamma}$ and $\bm{\theta}$ in the butterfly would be uniquely derived. Therefore, all the $N/2$ butterflies in the $N$-state trellis can be classified to $2^R$ (denoted by $N_c$) groups. The groups are distinguished by $\bm{\alpha}$, which means that butterflies in the same group have the same branch metrics at one stage.
As a result, for the $N / N_c$ states in the same group, only four branch metrics need to be calculated, to update the $N / N_c$ path metrics. Thus, the total computation of branch metrics for all the ACS operations at one stage can be calculated as $2^{R+2}$. For the widely used convolutional codes which have $R=2$ and $K=5,7,9$, or $R=3$ and $K=7,9$, the forward ACS operations can be accelerated due to lower computation of branch metrics than state-based or butterfly-based parallelism scheme ($2^{R+2} < 2^K$).

\section{Framework of Kernels and Memory Organization on GPU}

\subsection{Kernel Execution and Thread Mapping Strategies}
%
%

\renewcommand\arraystretch{1.1}
\begin{table}[bp]
\centering
\caption{Thread Dimensions and Execution Parallelism of Two Kernels}
\label{Tab_Parallelism}
\begin{tabular}{ccccc}
\hline
\multirow{2}{*}{Kernel} & \multicolumn{2}{c}{Thread dimension} & \multicolumn{2}{c}{Parallelism}\\
\cline{2-5}
 & BlockDim & \multicolumn{1}{c|}{ThreadDim} & Inter-frame & Intra-frame \\
\hline
K1 & $N_{bl}$ & \multicolumn{1}{c|}{$32N_c$} & $32N_{bl}$ & $N_c$ \\
K2 & $N_{bl}/N_c$ & \multicolumn{1}{c|}{$32N_c$} & $32N_{bl}$ & $1$ \\
\hline
\end{tabular}
\end{table}


In our GPU-based implementation, two individual kernels K1 and K2 with different thread dimensions are initiated. K1 finishes the forward computing, followed by K2 which carries out the traceback and decoding procedures. To describe the thread organizations in kernels, blockDim and threadDim are used to represent the number of threadblocks and the number of threads in each threadblock. In K1, the group-based parallel execution mode is employed. For the forward computing of a PB, all the $N$ states will be sorted to $N_c$ groups using the given criteria. Then for each group, a thread is dispatched to calculate four (or two in special) branch metrics to update all the path metrics and survivor paths at each stage. Thus, $N_c$ threads are required to build a virtual processor in K1. Considering that 32 CUDA threads are managed cooperatively in batches called a warp, a threadblock in K1 is regulated to accommodate 32 virtual processors. That means the threadDim of K1 is $N_c$ times the warp size.


\begin{algorithm}[tb]
\caption{Parallel block-based Viterbi decoding algorithm}
\label{Alg_PBVD}

\begin{algorithmic}[1]
\KERNELONE
\textbf{Forward procedure}
\FOR {thread block $b=0$ to $N_{bl} -1$, warp $w=0$ to $N_c-1$ and thread $t=0$ to $31$ \textbf{parallel}}
\FOR {stage $s=0$ to $D+2L-1$}
\STATE $sp = 0$, $tid= b \times 32 + t$;
\STATE Load input symbol and calculate four branch metrics;
\FOR {\textbf{all} $j \in Group(w)$}
\STATE Load: $pm_1 = {\rm PM}[2j][t]$, $pm_2 = {\rm PM}[2j+1][t]$;
\STATE $reg[j] = min( pm_{1} + BM_{\bm{\alpha}}, pm_{2} + BM_{\bm{\gamma}} )$;
\STATE take a bitwise record in $sp$ for state $j$;
\STATE $reg[j+2^{K-2}] = min( pm_{1} + BM_{\bm{\alpha}}, pm_{2} + BM_{\bm{\gamma}} )$;
\STATE take a bitwise record in $sp$ for state $j+2^{K-2}$;
\ENDFOR
\STATE Store: ${\rm PM[\ast][t]} = reg[\ast]$, ${\rm SP}[s][w][tid] = sp$;
\ENDFOR
\ENDFOR

\KERNELTWO
\textbf{Backward procedure}
\FOR {thread block $b=0$ to $N_{bl} / N_c - 1$, warp $w=0$ to $N_c-1$ and thread $t=0$ to $31$ \textbf{parallel}}
\STATE $i=j=g=state=0$, $tid=b \times N_c \times 32 + w \times 32 + t$;
\FOR {stage $s=D+2L-1$ to $L$}
\STATE Obtain $i$ by $state$ from lookup tables;
\FOR {$g=0$ to $N_c-1$}
\STATE Load ${\rm SP}[s][g][tid]$ and store into $sp$;
\ENDFOR
\IF {$s \leq D+L-1$}
\STATE Output decoded bit: $(state>>(K-2)) \& 0x01$;
\ENDIF
\STATE $j = state \% 2^{K-2}$, $sp = (sp >> i) \& 0x01$;
\STATE $state = 2 \times j + sp$;
\ENDFOR
\ENDFOR
\end{algorithmic}

\end{algorithm}

In the second kernel K2, as the backward procedure is a completely serial processing that can not be executed in parallel, only one thread is enough to constitute the virtual processor in K2. For convenience narration, we let the threadDim of K2 equal to K1, so that each threadblock in K2 contains $32 \times N_c$ virtual processors. If we allocate $N_{bl}$ threadblocks in K1, the total number of PBs $N_t$ should be equal to $32 \times N_{bl}$. Thus, to handle the $N_t$ PBs simultaneously, $N_{bl} / N_c$ threadblocks should be allocated in K2. Inter-frame parallelism and intra-frame parallelism are introduced to indicate the number of virtual processors in each kernel and the number of threads each virtual processor contains, respectively. Table \ref{Tab_Parallelism} gives a summary about the thread dimensions and execution parallelism of K1 and K2.

\begin{figure}[tb] 
\centering
\includegraphics[width=3.4in]{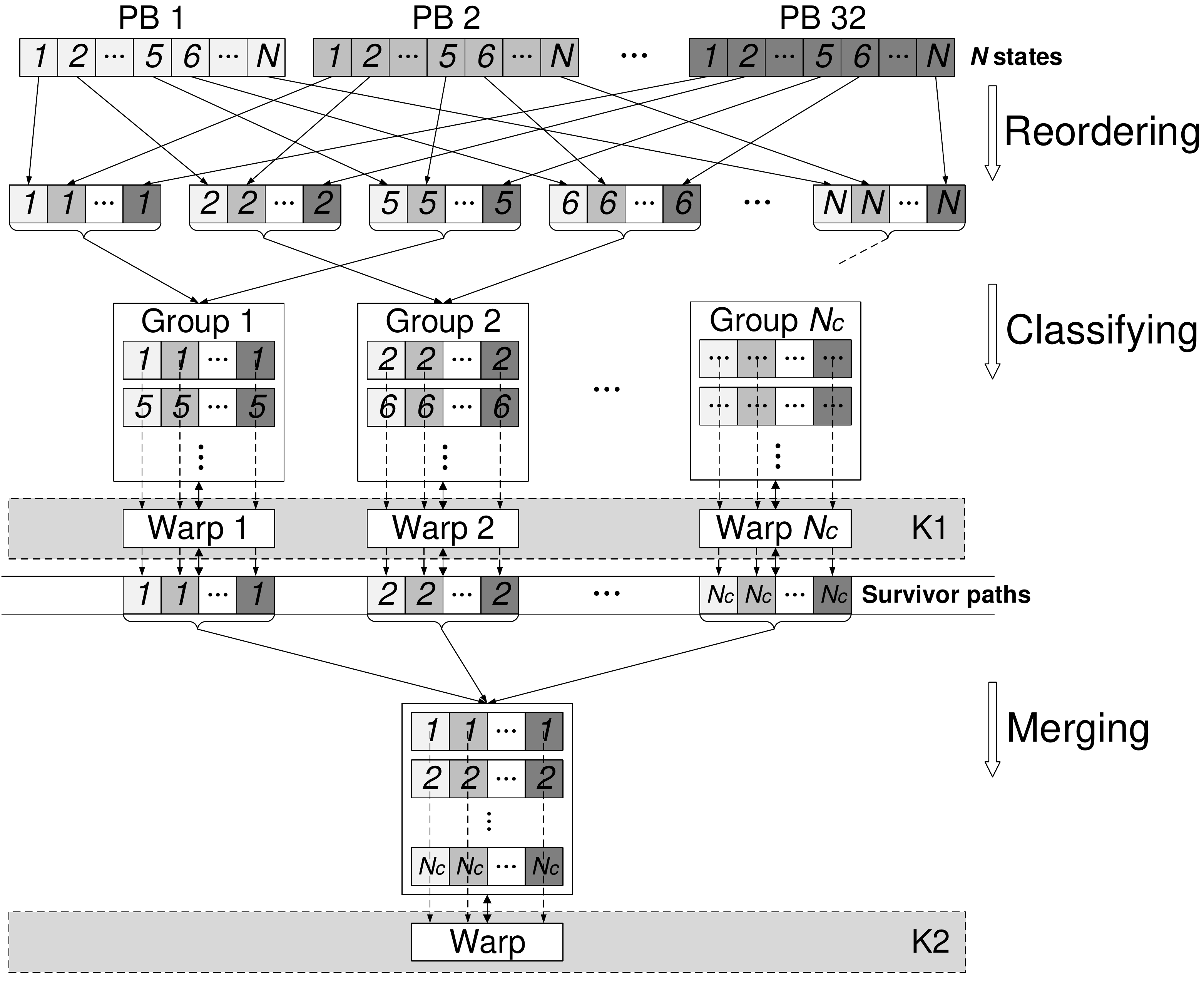}
\caption{The diagram of coalesced memory accesses for survivor paths.}
\label{Fig_Coalesced}
\end{figure}

\subsection{Memory Organization for Various Information}
In the parallel block-based Viterbi decoder, there are several kinds of data information: (i) the input/output data streams, which can only be stored in the off-chip global memory as they need to be exchanged from the host machine; (ii) the cumulative path metrics and the branch metrics, which are only updated in the forward procedure, so on-chip register resources and shared memory can be used under the conditions of enough capacity; (iii) the survivor paths, which are generated in forward procedure and fetched in decoding phase, that they can only be placed in global memory and designed to meet the alignment requirement for coalesced memory access of two individual kernels.

It is a challenge to design a suitable data structure for the survivor paths due to the different intra-frame parallelism in K1 and K2. Once the coalesced memory access is satisfied in one of the two kernels, the memory transactions in the other kernel would face to horrible inefficiencies.
To solve this inconsistency, an optimized construction is exploited in Fig.\ref{Fig_Coalesced}. At the current stage, states from 32 PBs are gathered and reordered. All states would be collected to $N_c$ groups followed by the state classification criteria. These $N_c$ groups of states are mapped to different warps allocated in K1. Inside a group, $N/N_c$ states from the same PB are processed in order by the same thread, which means threads with the same threadIdx.x from these $N_c$ warps make up a virtual processor in K1. As the survivor path is a record of selected forward path which can be presented by bit data (for example, bit 0 denotes the upper branch, and bit 1 denotes the lower branch), these $N/N_c$ results can be stored by bit in a same unit as:
\begin{equation}\label{Eq_bit}
{\rm SP}[ x ][ y ][ z ] = \underbrace {1101 \cdot  \cdot  \cdot 01}_{(N/{N_c})bits} \notag
\end{equation}
As a result, the survivor paths should be allocated as ${\rm SP}[D+2L][N_c][N_t]$
to ensure coalesced access for contiguous PBs inside a warp. For each backward stage in K2, $N_c$ individual results are merged because only one warp is needed for the backward phase of these 32 PBs. For a single thread in this warp, all survivor path messages from a PB are loaded with $N_c$ memory requests, but all in the form of aligned transaction. After all, the memory requests in both K1 and K2 are managed without duplicate transactions and extra time overhead.

The shared memory are allocated based on thread blocks and threads with the same threadIdx.x in different warps need to swap data to jointly accomplish the forward phase for a PB. To avoid the bank conflict in shared memory transactions, the data structure should be devised as ${\rm PM}[N][32]$
to ensure that the accesses for path metrics with the same state id are aligned and fall into individual shared memory banks. As a result, for each shared memory store/load instruction, no transaction for the same request replays and maximum bandwidth utilization is reached.
Remarkably,
additional registers are necessary as the temporary places to store the updated results for path metrics, and shared memory store transactions would not be carried out until all the calculations at a stage are finished.

\subsection{Asynchronous Data Transfer and Throughput Analysis}

The time overhead of data transfer between host and device should be taken into account when evaluating a GPU-based decoder. CUDA supports asynchronous streams technique to achieve the overlap for data transfer tasks and kernel launches in different streams. The decoder should activate a suitable number of CUDA streams and arrange tasks to the idle streams consecutively to ensure the high occupancy of the GPU device.

For our GPU-based Viterbi decoder, the H2D messages are blocked input data streams and D2H messages are decoded bits. A kernel throughput $S_k$ is introduced to evaluate the kernel execution efficiency and it can be obtained by $\frac{D \times N_{t}}{T_k}$,
where $T_k$ is kernel execution time. For the H2D data transfer, a parameter $U_1$ is defined to indicate the number of bytes for an input symbol storage. Similarly, a parameter $U_2$ is defined to indicate the number of bytes for the storage of a decoded bit in D2H data transfer. Thus, the time cost of H2D and D2H transfer can be calculated by: $T_{\rm H2D} = \frac{(D+2L) \times N_{t} \times U_1}{B}$ and $T_{\rm D2H} = \frac{D \times N_{t} \times U_2}{B}$,
where $B$ denotes the PCI-E bandwidth. To hide data transfer latency, $N_s$ CUDA streams can be allocated (in each stream, $N_t$ parallel blocks are arranged). 
Ideally, all the data transfer batches can be completely hidden by the kernel executions, besides the first H2D batch and the last D2H batch. Thus, the decoding throughput can be approximately calculated by :
\begin{align}\label{Eq_TP}
{\rm T/P} &\approx \frac{D \times N_t \times N_s}{T_{\rm H2D} + \sum{T_k} + T_{\rm D2H}} \notag\\
    &\approx \frac{B \times N_s}{(1+2L/D) \times U_1 + N_s/S_k + U_2}
\end{align}
Notice that the approximation $\sum{T_k} \approx N_s \times T_k$ can be used, though the concurrent kernel execution (CKE) technique or the Hyper-Q technique in CUDA may be applied.


\renewcommand\arraystretch{0.9}
\begin{table}[bp]
\centering
\caption{Classification of states for a (2, 1, 7) convolutional code}
\label{Tab_Group}
\begin{tabular}{|c|c|c|c|c|c|}
\hline
Group & $\bm{\alpha}$ & $\bm{\beta}$ & $\bm{\gamma}$ & $\bm{\theta}$ & Index of states\\
\hline
\multirow{2}{*}{0} & \multirow{2}{*}{00} & \multirow{2}{*}{11} & \multirow{2}{*}{11} & \multirow{2}{*}{00} &  \multicolumn{1}{l|}{0, 1, 4, 5, 24, 25, 28, 29, 42, 43}\\
& & & & & \multicolumn{1}{l|}{46, 47, 50, 51, 54, 55}\\\hline
\multirow{2}{*}{1} & \multirow{2}{*}{01} & \multirow{2}{*}{10} & \multirow{2}{*}{10} & \multirow{2}{*}{01} & \multicolumn{1}{l|}{2, 3, 6, 7, 26, 27, 30, 31, 40, 41}\\
& & & & & \multicolumn{1}{l|}{44, 45, 48, 49, 52, 53}\\\hline
\multirow{2}{*}{2} & \multirow{2}{*}{11} & \multirow{2}{*}{00} & \multirow{2}{*}{00} & \multirow{2}{*}{11} & \multicolumn{1}{l|}{8, 9, 12, 13, 16, 17, 20, 21, 34}\\
& & & & & \multicolumn{1}{l|}{35, 38, 39, 58, 59, 62, 63}\\\hline
\multirow{2}{*}{3} & \multirow{2}{*}{10} & \multirow{2}{*}{01} & \multirow{2}{*}{01} & \multirow{2}{*}{10} & \multicolumn{1}{l|}{10, 11, 14, 15, 18, 19, 22, 23, 32}\\
& & & & & \multicolumn{1}{l|}{33, 36, 37, 56, 57, 60, 61}\\
\hline
\end{tabular}
\end{table}


To improve the decoding throughput, one way is to make the kernels operate efficiently by the approaches in above sections, to achieve a higher $S_k$. Another way is to develop suitable methods of the storage for input/output messages, to reduce $U_1$ and $U_2$. For a soft-decision decoding over the AWGN channel, received symbols should be converted to soft messages and stored by several bits.
In fact, a $q$-bit fixed-point quantization scheme can be designed and $\lfloor 32/q \rfloor$ messages can be packed and stored into a same integer unit. As a result, the value $U_1$ decreases from $4R$ to $4R/\lfloor 32/q \rfloor$. For the storage of decoded bits,
we can use a similar packing scheme to store each decoded bit by bitwise operations. In this case, a character type can store 8 individual decoded bits that reduce $U_2$ to $1/8$.



\renewcommand\arraystretch{1.0}
\begin{table*}[htbp]
\centering
\caption{Time consumption and throughput of original and optimized decoder under different devices and various Parallelism}
\label{Tab_R2}
\begin{tabular}{cccccccc|ccccccc}
\hline
\multirow{2}{*}{Device} & \multirow{2}{*}{$N_{bl}$} & \multirow{2}{*}{$N_t$} & \multicolumn{5}{c|}{Original results} & \multicolumn{7}{c}{Optimized results}\\
\cline{4-15}
& & & $T_k$ & $T_{\rm H2D}$ & $T_{\rm D2H}$ & $S_k$ & T/P(1S) & $T_{k1}$ & $T_{k2}$ & $T_{\rm H2D}$ & $T_{\rm D2H}$ & $S_k$ & T/P(1S) & T/P(3S)\\
\hline
\multirow{5}{*}{GTX580} & 64 & 2048 & 2.914 & 1.532 & 0.636 & 359.8 & 181.5 & 1.443 & 0.611 & 0.377 & 0.023 & 509.5 & 403.4 & 508.3\\
& 128 & 4096 & 5.811 & 2.968 & 1.280 & 362.9 & 185.4 & 3.046 & 0.859 & 0.747 & 0.043 & 571.4 & 446.4 & 547.7\\
& 192 & 6144 & 8.514 & 4.506 & 1.969 & 368.0 & 189.1 & 4.050 & 1.232 & 1.155 & 0.063 & 594.5 & 472.2 & 571.0\\
& 256 & 8192 & 11.361 & 5.986 & 2.556 & 368.2 & 189.3 & 5.250 & 1.456 & 1.571 & 0.082 & 628.7 & 498.4 & 590.0\\
& 320 & 10240 & 14.224 & 7.502 & 3.192 & 369.6 & 189.4 & 6.513 & 1.807 & 1.893 & 0.101 & 641.8 & 504.9 & 598.3\\ \hline
\multirow{5}{*}{GTX980} & 64 & 2048 & 1.681 & 0.865 & 0.325 & 620.6 & 294.7 & 0.591 & 0.377 & 0.261 & 0.012 & 1082.5 & 764.9 & 1243.5\\
& 128 & 4096 & 3.232 & 1.771 & 0.652 & 647.1 & 298.6 & 0.840 & 0.386 & 0.454 & 0.023 & 1575.4 & 1051.4 & 1623.7\\
& 192 & 6144 & 4.831 & 2.684 & 0.981 & 650.8 & 304.9 & 1.172 & 0.392 & 0.678 & 0.032 & 2005.2 & 1253.0 & 1767.5\\
& 256 & 8192 & 6.436 & 3.613 & 1.333 & 652.3 & 308.8 & 1.568 & 0.414 & 0.896 & 0.042 & 2116.8 & 1290.6 & 1785.2\\
& 320 & 10240 & 8.034 & 4.334 & 1.657 & 652.5 & 309.1 & 1.899 & 0.523 & 1.102 & 0.052 & 2122.7 & 1324.7 & 1802.5\\
\hline
\end{tabular}
\begin{tablenotes}
    \item $T_k$, $T_{\rm H2D}$ and $T_{\rm D2H}$ are in ms. $S_k$ and T/P are in Mbps.
   \end{tablenotes}
\end{table*}

\section{Experimental Results and Discussions}
The experimentations are carried out on Intel i7-4790k platform with NVIDIA GTX580 (1544MHz,  512 CUDA cores, and PCI-E 2.0 supported) and Nvidia GTX980 (1126MHz, 2048 CUDA cores, and PCI-E 3.0 supported). The programs
are complied with GCC 4.8.2 and CUDA 6.5.

\begin{figure}[tb] 
\centering
\includegraphics[width=3.0in]{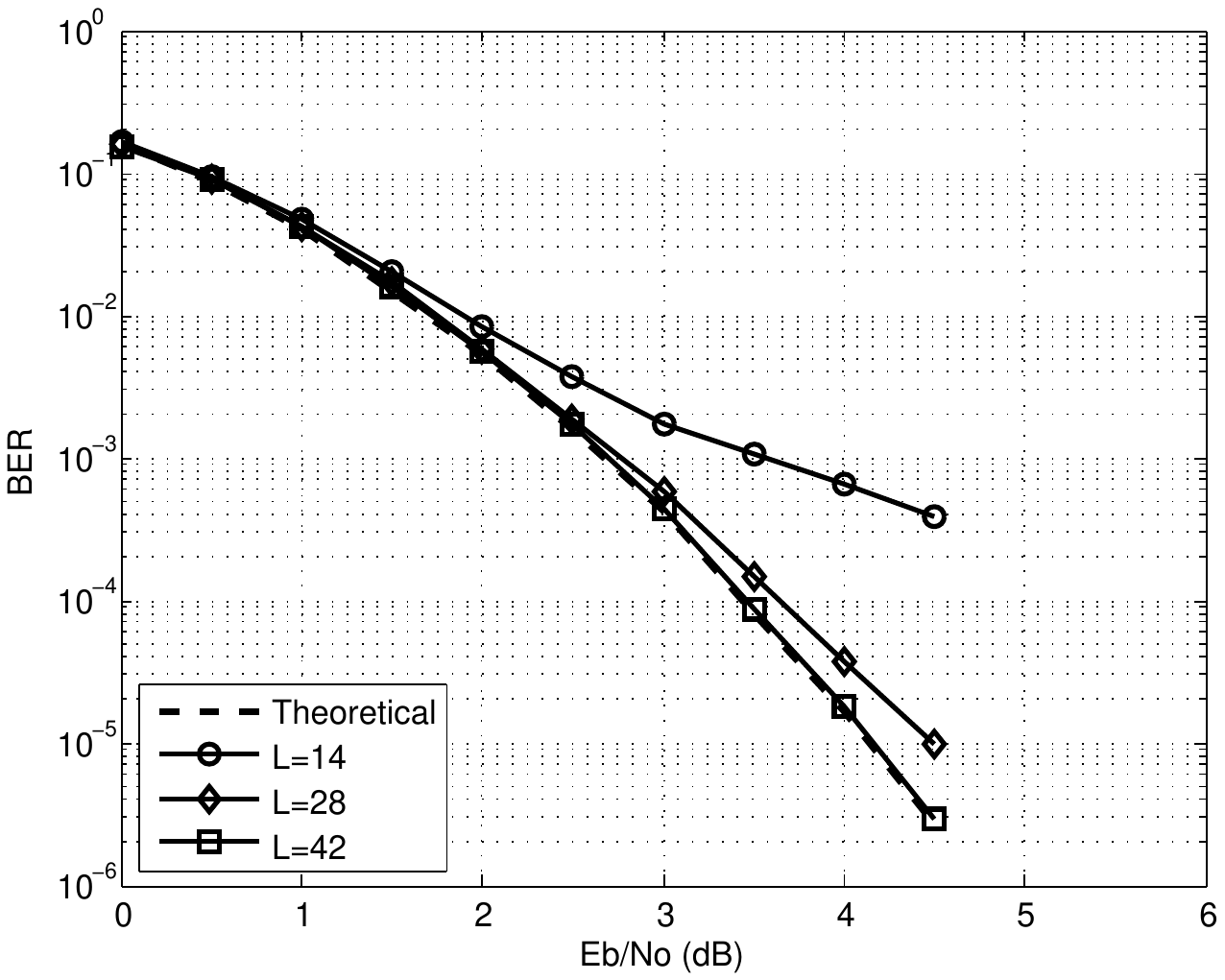}
\caption{BER performance of the (2,1,7) convolutional code. ($D=512$, 8-bit quantization)}
\label{Fig_BER}
\end{figure}

A (2,1,7) convolutional code with generator polynomials $\textbf{\emph{g}}^{(1)}=[1111001]$ and $\textbf{\emph{g}}^{(2)}=[1011011]$ is chosen from CCSDS standard \cite{CCSDS} for convenient comparison with other works. As the code rate is 1/2, the 64 states can be divided into $2^2=4$ groups using the given classification methods, and the result is shown in Table \ref{Tab_Group}. The BER performance under AWGN channel for various $L$ are presented in Fig.\ref{Fig_BER} ($D$ is fixed to 512, which is an less important factor). It is shown that as $L$ rises to 42, which is about 6 times the constraint length, the BER result is approximate to the theoretical performance. Actually, in the proposed decoder, larger $L$ results in better error correction performance, but too large $L$ can cut down the decoding throughput. Thus, $D=512$ and $L=42$ are selected for the parallel block in the following tests.

To demonstrate the improvements by using the proposed strategies and methods, experimental results of both the original decoder and the optimized decoder are given for comparison in Table \ref{Tab_R2}, including the kernel execution times, the data transfer times, the kernel throughput and the decoding throughput. The proposed decoder is operated on different GPU devices with various numbers of $N_{bl}$ and $N_t$. The original parallel block-based Viterbi decoder launches only one kernel to finish the whole decoding procedure. 32-bit float-point quantization is used for the input soft messages, and decoded bits are stored in integers. In the optimized decoder, two kernels with different number of threads are launched and execution times $T_{k1}$ and $T_{k2}$ are recorded individually. It can be seen that the total execution times are reduced significantly by at least 40\%, which results in an improvement of kernel throughput $S_k$. Input messages are quantized to 8-bit, which are stored using the packing scheme, and bitwise storage is designed for decoded bits. As a result, the H2D/D2H data transfer sizes are both cut down and $T_{\rm H2D}$/$T_{\rm H2D}$ are greatly shorted to improve the decoding throughput (T/P). To hide data transfer latency, asynchronous transfer technique is adopted and throughput results with three CUDA streams (3S) are presented. By comparing with the performance under the synchronous mode which only uses one CUDA stream (1S), it shows that the more powerful the GPU is, the more efficient overlap and more throughput improvement become. Futhermore, as the increase in the number of concurrently executed parallel blocks $N_t$, the GPU will finally run at full capacity and the decoder will reach the peak throughput.

Table \ref{Tab_R3} shows the decoding throughput comparison between our work and existing works on various GPU platforms, which are all for convolutional codes with code rate 1/2 and constraint length 7. A metric named TNDC (Throughput under Normalized Decoding Cost) introduced in \cite{TNDC} is provided in order to make fair comparison. As the normalized results show, the proposed decoder achieves about 1.5$\sim$9.2 times speedup compared with the existing GPU-based implementations.



In addition, compared with the existing fastest x86-CPU work \cite{CPU2010}, which runs a 64-state VA decoder on the Intel Core 2 Extreme X9650 (4 cores, 3.0GHz) at the speed of 60Mbps, our results show significant throughput advantages. Compared with the newest results on FPGA platforms, e.g., 865Mbps for a 64-state VA decoder on Stratix III 340 (216MHz) \cite{FPGA2014} and 10Gbs for a 32-state VA decoder on Xilinx Virtex 7 XC7VX690T-2 \cite{FPGA2015}, our results reach a comparable speed, and the good scalability and compatibility make it easy to transplant our decoder onto future powerful GPU devices to achieve higher performance.

\renewcommand\arraystretch{1.1}
\begin{table}[htb]
\centering
\caption{Decoding throughput comparison with existing works}
\label{Tab_R3}
\begin{tabular}{ccccc}
\hline
Work & Device & T/P(Mbps) & TNDC & Speedup\\
\hline
\cite{SDR2011} & GTX275 & 28.7 & $ \approx $0.085 & $\times$9.20\\
\cite{TVDA2011} & 8800GTX & 29.4 & $ \approx $0.170 & $\times$4.60\\
\cite{TVDA_WCNC2013} & GTX580 & 67.1 & $ \approx $0.085 & $\times$9.20\\
\cite{SDR2010} & 9800GTX & 90.8 & $ \approx $0.420 & $\times$1.86\\
\cite{OPENCL2014} & HD7970 & 391.5 & $ \approx $0.207 & $\times$3.78\\
\multirow{2}{*}{\cite{TVDA_2014}} & Tesla C2050 & 240.9 & $ \approx $0.468 & $\times$1.67\\ 
& GTX580 & 404.7 & $ \approx $0.512 & $\times$1.53\\\hline 
\multirow{2}{*}{This work} & GTX580 & 598.3 & $ \approx $0.757 & $\times$1.03\\
& GTX980 & 1802.5 & $ \approx $0.782 & $\times$1.00\\ 
\hline
\end{tabular}
\end{table}

\section{Conclusion}
This paper introduces a parallel block-based Viterbi decoder. The data stream is divided to a series of parallel blocks for concurrently decoding. Implementation on GPU uses two individual kernels mapping to two decoding phases, and optimized parallelism inside kernels are presented, which are based on the proposed state classification criteria. Aiming to accelerate the decoding, appropriate GPU memory and data structure are developed for intermediate messages. Storage for input/output data are designed and multiple CUDA streams are used to reduce the overhead of data transfer. Experimental results show that proposed GPU-based decoder achieves about 1.5 times speedup than the existing fastest work on GPU. The proposed decoding architecture can be used in the software-defined radio systems, as a flexible Viterbi decoding unit with strong reconfigurable ability.


\section*{Acknowledgment}

This work was supported by the National Natural Science Foundation of China (91438116).



%
%
%

\bibliographystyle{IEEEtran}

\bibliography{IEEEabrv,mybibfile}

\end{document}